\documentclass[%
 preprint,
superscriptaddress,
 amsmath,amssymb,
 aps,
prb,
]{revtex4-2}

\usepackage[utf8] {inputenc}
\usepackage[english] {babel}
\usepackage{graphicx}
\usepackage{dcolumn}
\usepackage{bm}
\usepackage{mathtools}
\usepackage{float}
\usepackage{multirow}
\usepackage{amsmath}

\usepackage{lmodern}
\usepackage{textcomp}
\usepackage{t1enc}

\usepackage{graphicx}
\usepackage{ulem}
\usepackage{setspace}

\usepackage[toc, page]{appendix}
\usepackage{standalone}

\usepackage{verbatim}

\usepackage{natbib}

\usepackage{comment}

\newlength{\depthofsumsign}
\newlength{\totalheightofsumsign}
\newlength{\heightanddepthofargument}

\newcommand{\nsum}[1][1.4]{
    \mathop{%
        \raisebox
            {-#1\depthofsumsign+1\depthofsumsign}
            {\scalebox
                {#1}
                {$\displaystyle\sum$}%
            }
    }
}

\makeatletter
\newcommand*{\DivideLengths}[2]{%
  \strip@pt\dimexpr\number\numexpr\number\dimexpr#1\relax*65536/\number\dimexpr#2\relax\relax sp\relax
}
\makeatother

\begin{document}


\title{Identification of silicon vacancy-related electron paramagnetic resonance centers in 4H SiC}

\author{A. Cs\'or\'e}
\affiliation{%
 Department of Atomic Physics, Budapest University of Technology and Economics, Budafoki \'ut 8., H-1111, Budapest, Hungary}
 
\author{N. T. Son}
\affiliation{%
Department of Physics, Chemistry and Biology, Link\"oping University, SE-58183 Link\"oping, Sweden}

\author{A. Gali}
 \affiliation{%
 Department of Atomic Physics, Budapest University of Technology and Economics, Budafoki \'ut 8., H-1111, Budapest, Hungary}
\affiliation{%
 Wigner Research Centre for Physics, PO. Box 49., Budapest H-1525, Hungary}

\date{\today}

\begin{abstract}
The negatively charged silicon vacancy [V$_\text{Si}(-)$] in silicon carbide (SiC) is a paramagnetic and optically active defect in hexagonal SiC. V$_\text{Si}(-)$ defect possesses $S = 3/2$ spin with long spin coherence time and can be optically manipulated even at room temperature. Recently, electron spin resonance signals have been observed besides the signals associated with the V$_\text{Si}(-)$ defects in the 4H polytype of SiC. The corresponding centers share akin properties to those of the V$_\text{Si}(-)$ defects and thus they may be promising candidates for quantum technology applications. However, the exact origin of the new signals is unknown. In this paper we report V$_\text{Si}(-)$-related pair defect models as possible candidates for the unknown centers. We determine the corresponding electronic structures and magneto-optical properties as obtained by density functional theory (DFT) calculations. We propose models for the recently observed electron paramagnetic resonance centers with predicting their optical signals for identification in future experiments. 
\end{abstract}

\pacs{Valid PACS appear here}
\maketitle

\section{\label{intro}Introduction}

Paramagnetic point defects in solids have attracted a great attention as they can act as quantum bits (qubits) and single photon sources which are the building blocks of quantum technology applications. In particular, point defects embedded in diamond and silicon carbide (SiC) are leading candidates as these hosts provide wide band gaps allowing large separation of the deep defect levels.

In particular, the negatively charged silicon vacancy [V$_\text{Si}(-)$] in silicon carbide (SiC) is one of the most studied defects in hexagonal SiC polytypes. V$_\text{Si}(-)$ defects exhibit $S = 3/2$ spin state as observed in electron paramagnetic resonance (EPR) experiments~\cite{WimbauerPRB1997}. Furthermore, at single defect level the ground state exhibits long spin coherence time and can be optically manipulated even at room temperature~\cite{Widmann2015NatMat}. The corresponding optical emission of all V$_\text{Si}(-)$ defects fall into the near-infrared (NIR) region with zero-phonon lines (ZPLs) of 1.438~eV and 1.352~eV in the 4H polytype denoted as V$_1$ and V$_2$, respectively~\cite{SormanPRB2000, NagyNatCom2019}. These outstanding properties make V$_\text{Si}(-)$ defects highly promising qubits~\cite{KrausNatPhys2014, BaranovPRB2011, RiedePRL2012, SoltamovPRL2012, FuchsNatComm2015, FuchsSciRep2013, CarterPRB2015, SoykalPRB2016, Widmann2015NatMat, SiminPRB2017, RadulaskiNanoLett2017, NagyPRapplied2018} in ultrasensitive nanosensor applications such as magnetometry~\cite{SangYunPRB2015, KrausSciRep2014, Simin2016PRX, SiminPRApplied2015, NiethammerPRApplied2016, CochraneSciRep2016} and thermometry~\cite{AnisimovSciRep2016, KrausSciRep2014}.

A rich set of data has been accumulated in the past decades from magnetic resonance experiments on V$_\text{Si}(-)$ in 4H SiC, however many details are not resolved yet. Owing to the lattice structure of the 4H polytype, two different V$_\text{Si}(-)$ defects can be formed: one is located at a hexagonal lattice site labeled as V$_\text{Si}(-)$-$h$ and V$_\text{Si}(-)$-$k$ residing at a quasicubic site. In this way two different photoluminescence (PL) and EPR signals are expected, however, four V$_\text{Si}(-)$-related EPR signals have been observed labeled as T$_\text{1va}$, T$_\text{2va}$, T$_\text{1vb}$, T$_\text{2vb}$~\cite{SormanPRB2000, MizuochiPRB2002, MizuochiPRB2003, MizuochiPRB2005, JanzenPhysB2009} and further two denoted as R$_1$ and R$_2$ which have been recently attributed to this family~\cite{SonIOP2019}. Common properties of these centers are the $S = 3/2$ ground state, the C$_\text{3v}$ defect symmetry, the same isotropic $g$-value of 2.0029~\cite{SonIOP2019} and the relatively small zero-field splittings (ZFS) characterized by the $D$ tensor (cf.~Table~\ref{tab:expDconst}). Alternative identifications have been reported for the T$_\text{V}$ signals suggesting V$_\text{Si}(0)$~\cite{BardelebenPRB2000} and V$_\text{Si}(-)$-V$_\text{C}(0)$ complex~\cite{KrausNatPhys2014, SoltamovPRL2015} defect models where V$_\text{C}(0)$ is located at the third and seventh neighbor along the crystal axis ($c$-axis)~\cite{MizuochiPRB2002,  MizuochiPRB2005}. Nevertheless, both models have been nullified by recent density functional theory (DFT) calculations~\cite{IvadyPRB2017_1, DavidssonAPL2019} assigning T$_\text{1va}$ to V$_\text{Si}(-)$-$h$ and T$_\text{2va}$ to V$_\text{Si}(-)$-$k$, but the remaining four signals are remained to be unidentified.

In this work, we employ \textit{ab initio} DFT calculations in order to identify the V$_\text{Si}(-)$-related EPR centers. To this end, we establish defects models that might act as the origin of the reported EPR signals. These models are introduced in Sec.~\ref{sec:defmod}. The applied methodology is described in Sec.~\ref{sec:method}, particularly, we summarize the utilized computational techniques in Subsec.~\ref{subsec:compmethod}; we describe the formulation and the derivation of formation and binding energies in Subsec.~\ref{subsec:formmethod} and parameters of ZFS are introduced in Subsec.~\ref{subsec:ZFSmethod}. We present our results in Sec.~\ref{sec:resdisc}, in particular, the corresponding electronic structures in Subsec.~\ref{subsec:elstruct}, defect formation in Subsec.~\ref{subsec:defform} and the calculated $D$ constants and ZPLs for all defect models in Subsec.~\ref{subsec:magneto}. Although only EPR centers exhibiting C$_\text{3v}$ symmetry have been reported, we dedicate Subsec.~\ref{subsec:EPR_C1h} to disuss the difficulties in resolving EPR spectrum of defects with C$_\text{1h}$ symmetry. We conclude our work in Sec.~\ref{sec:sum}.


\section{\label{sec:defmod}Defect models}

Since the four unidentified EPR centers share closely related properties to those of the isolated V$_\text{Si}(-)$ defects, they may introduce similar ground state electronic structure with $S = 3/2$ spin state. However the corresponding spin densities may be slightly modified yielding different $D$ constants (cf.~Table~\ref{tab:expDconst}). Consequently, defect complexes consiting V$_\text{Si}(-)$ and another single defect denoted as $X$ may be suitable candidates for this role where $X$ is expected to act perturbatively on the electronic structure of V$_\text{Si}(-)$. In order to preserve the C$_\text{3v}$ symmetry reported for all EPR centers~\cite{SonIOP2019} $X$ should be located along the $c$-axis establishing axial V$_\text{Si}(-)$-$X$ complexes. Here we note that such defect models have already been proposed~\cite{MizuochiPRB2002, MizuochiPRB2005} following similar arguments. In particular, the V$_\text{Si}(-)$-V$_\text{C}(0)$ defect complexes comprising distant V$_\text{Si}(-)$ and V$_\text{C}(0)$ defects along the $c$-axis have been suggested~\cite{KrausNatPhys2014, SoltamovPRL2015} to be promising candidates, however the corresponding electronic structure forms $S = 1/2$ ground state as revealed by earlier DFT calculations~\cite{IvadyPRB2017_1}.

\begin{figure} [t]
\centering
\includegraphics[width=0.35\textwidth]{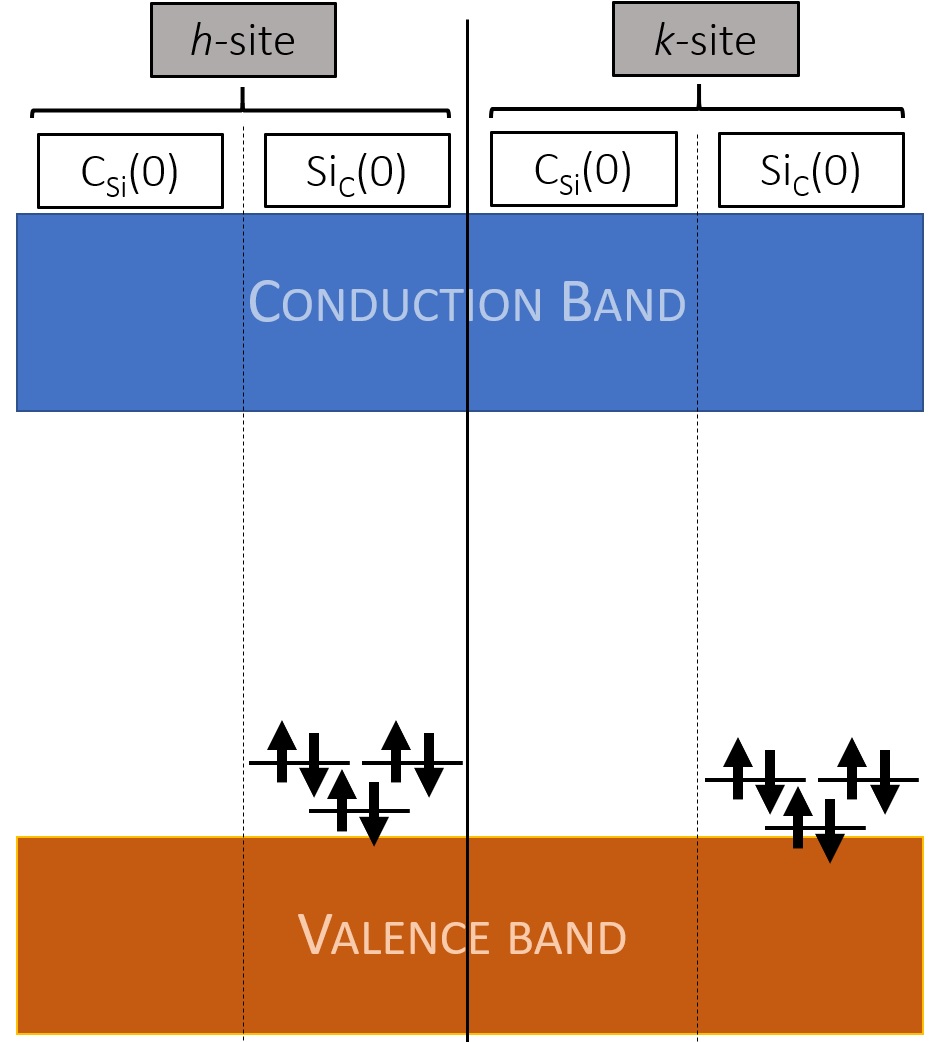}
\caption{Ingap Kohn-Sham levels introduced by antisites to the band gap of 4H SiC situated at both $h$ and $k$ sites. Energy levels for Si$_\text{C}(0)$ are slightly lower for Si$_\text{C}(0)$-$k$ than for Si$_\text{C}(0)$-$h$, while no states are located in the band gap neither for C$_\text{Si}(0)$-$h$ nor for C$_\text{Si}(0)$-$k$.}
\vspace{0 pt}
\label{fig:antielstruct}
\end{figure}

\begin{figure*} [t]
\centering
\includegraphics[width=\textwidth]{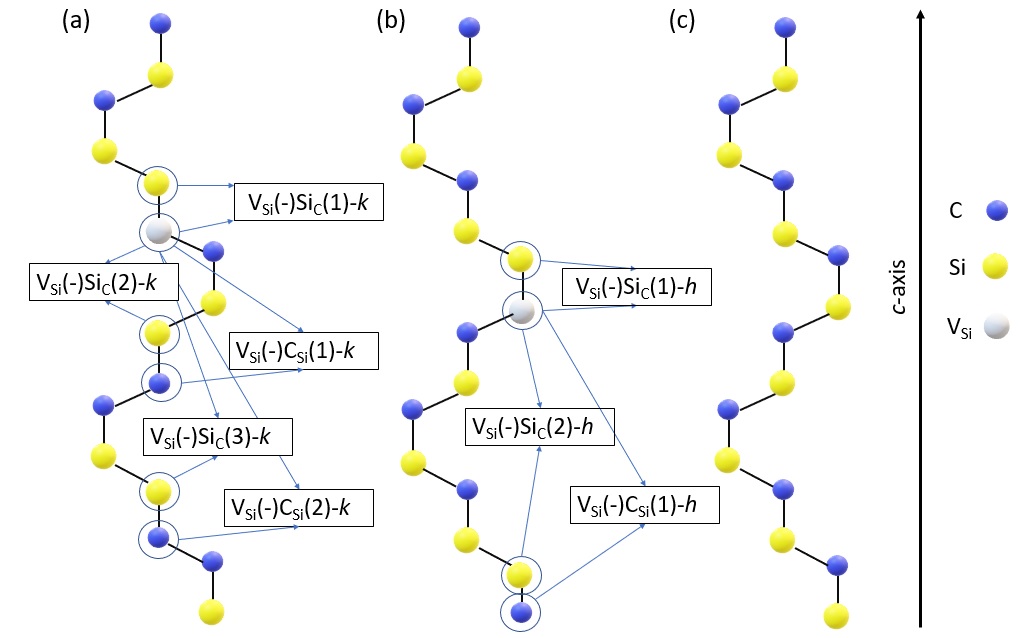}
\caption{Defect models of axial V$_\text{Si}(-)$-$X$ defect complexes, where $X=\{$C$_\text{Si}(0)$, Si$_\text{C}(0)\}$. Neighboring order between V$_\text{Si}(-)$ and $X$ are indicated in brackets. The possible (a) V$_\text{Si}(-)$-$X$-$k$ and (b) V$_\text{Si}(-)$-$X$-$h$ pair defects along with the (c) perfect lattice are illustrated. Labels of atoms and the $c$-axis are indicated.}
\vspace{0 pt}
\label{fig:defstructs}
\end{figure*}

Previous EPR measurements were carried out on 4H SiC grown by chemical vapor deposition (CVD), which were low-doped with a residual N-doping concentration of $\approx$ 5 $\times$ 10$^{12}$~cm$^{-3}$ as estimated from the corresponding PL spectrum~\cite{IvanovJAP1996}. This suggests that $X$ should be an intrinsic defect. In this work we investigate defect models involving distant V$_\text{Si}(-)$-$X$ defect pairs coordinated axially in the 4H SiC lattice, where we assign $X$ to the neutral carbon or silicon antisite defects denoted as C$_\text{Si}(0)$ and Si$_\text{C}(0)$, respectively, as depicted in Fig.~\ref{fig:defstructs}. These assignments are supported by the fact that the intrinsic antisites of SiC introduce spinless electronic structures. In particular, Si$_\text{C}(0)$ introduces low-lying fully occupied states, i.e. an $a_1$ and an $e$ state to the band gap of 4H SiC, whereas C$_\text{Si}(0)$ is electrically inactive. This remarkable difference might be attributed to the smaller size of the C atom against Si atom allowing the creation of lower-energy bonds. 
The corresponding electronic structures are illustrated in Fig.~\ref{fig:antielstruct}. The idea behind these models is that the antisite defects would introduce relatively small perturbation to the neighbor V$_\text{Si}(-)$ defect, in particular, C$_\text{Si}$, so antisite defects would not alter the spin state of V$_\text{Si}(-)$ defect but would modify the corresponding $D$ constants because $D$ constants are sensitive to the strain caused by the neighbor antisite defect. As  V$_\text{Si}(-)$ defects are created by irradiation techniques, cascade process of vacancy formations may lead to the creation of nearby antisite defects which are stable and immobile intrinsic defects.
Due to the crystal structure of the 4H polytype, if V$_\text{Si}(-)$ is located at an $h$/$k$ site then $X$ resides also in an $h$/$k$ layer along the $c$-axis.

\section{\label{sec:method} Methodology}

\subsection{\label{subsec:compmethod}Computational methodology}

All investigated defects were modeled in a 576-atom 4H supercell while for $D$-constant and ZPL calculations a 1536-atom supercell was also used. Both supercells are sufficiently large enabling only $\Gamma$-point sampling of the Brillouin-zone to reach convergent wavefunctions. Calculations were carried out by using the spinpolarized Heyd-Scuseria-Ernzerhof (HSE06)~\cite{HeydJCP2003} hybrid functional and the computationally economical Perdew-Burke-Ernzerhof (PBE)~\cite{PerdewPRL1996} functional. In particular, the electronic structure is described by the HSE06 functional applied on 576-atom supercell, while for calculations of the magneto-optical properties we employ PBE functional on 1536-atom supercell. For defect formation energies we test both functionals: we report results calculated by HSE06 functional in Sec.~\ref{subsec:defform}, while PBE results for defect formation are reported in Appendix A. Kohn-Sham (KS) wavefunctions were expanded in plane wave basis set with the cutoff energy of 420~eV. In the calculations only valence electrons were treated explicitly, core-electrons were considered in the framework of projector augmented wave (PAW) method~\cite{BlochlPRB1994} as implemented in the Vienna Ab-Initio Simulation Package (VASP)~\cite{KressePRB1996}. Fully relaxed geometries were obtained by minimizing the forces between the ions falling below the threshold of 0.01~eV/\AA. Geometry relaxation revealed that formation of V$_\text{Si}(-)$ with the nearest neighbor Si$_\text{C}(0)$ is highly unlikely, instead they recombine yielding V$_\text{C}(-)$ at both $h$ and $k$ sites. Electronic structure of V$_\text{C}(-)$ is significantly differ from that of  V$_\text{Si}(-)$, particularly, it introduces $S = 1/2$ ground state and already identified EPR centers in 4H SiC~\cite{TrinhPRB2013}. Thus, in this context we exclude the defect model of the V$_\text{Si}(-)$ with Si$_\text{C}(0)$ located at the nearest neighbor C site.

Besides the close similarities between the observed EPR centers, their --- relatively small --- axial component $D$ constants slightly differ, while the orthorombic component $E$ constants of the corresponding ZFSs are zero for all defects due to the C$_\text{3v}$ symmetry exhibited by these centers. In this way, unambiguous identficiation may be achieved by comparing the experimental $D$ constants to those yielded by DFT calculations. To this end, we applied the house-built code as implemented by Iv\'ady \textit{et al.}~\cite{IvadyPRB2014ZFS} on PBE wavefunctions yielded by 576 and 1536 atom supercell calculations. Beside the EPR signature, the ZPLs may also be very helpful in defect identification. To determine the corresponding ZPLs we employed 576-atom supercells and calculated the excited state of all defect models employing $\Delta$SCF method~\cite{TozerPCCP2000, GaliPRL2009}.

 \subsection{\label{subsec:formmethod}Formation and binding energies}

The corresponding EPR lines were measured on electron irradiated 4H SiC samples~\cite{SonIOP2019}. Consequently, stability and thus concentration of the defect complexes is governed by the binding energy between V$_\text{Si}(-)$ and $X$. Binding energy ($E_\text{b}$) can be calculated as

\begin{equation}
E_\text{b}(E_\text{F}) = E_\text{form}^{\text{V}_\text{Si}(-)}(E_\text{F}) + E_\text{form}^{X}(E_\text{F}) - E_\text{form}^{\text{V}_\text{Si}(-)\text{-}X}(E_\text{F}),
\label{eq:binde}
\end{equation}
where $E_\text{form}^{\text{V}_\text{Si}(-)}(E_\text{F})$, $E_\text{form}^{X}(E_\text{F})$ and $ E_\text{form}^{\text{V}_\text{Si}(-)\text{-}X}(E_\text{F})$ are the formation energies of V$_\text{Si}(-)$, $X$ and $\text{V}_\text{Si}(-)\text{-}X$ as a function of the Fermi level ($E_\text{F}$), respectively. According to this definition, $E_\text{b}>0$ implies that the formation of $\text{V}_\text{Si}(-)\text{-}X$ complex is favorable. Defect formation energies in the $q$ charge state can be calculated as~\cite{AradiPRB2001}   
\begin{equation}
\begin{aligned}
E_{\text{form}}^q = E_{\text{tot}}^q - \frac{n_{\text{Si}} + n_{\text{C}}}{2}\mu_{\text{SiC}} - \frac{\mu_{\text{Si}}^\text{b} - \mu_{\text{C}}^\text{b} - \delta\mu}{2}(n_\text{Si} - n_\text{C})\\
+ q(E_{\text{F}}+E_\text{VBM}) + \Delta V(q),
\label{eq:formE}
\end{aligned}
\end{equation}
where $E_{\text{tot}}^q$ is the total energy of the defective system in the $q$ charge state, $\mu_{\text{Si}}^\text{b}, \mu_{\text{C}}^\text{b}$ are the chemical potentials of Si atom in bulk Si and C atom in diamond, respectively, $E_\text{VBM}$ represents the valence band edge and $\Delta V(q)$ stands for the charge correction term. To determine $\Delta V(q)$, we use the Freysoldt charge correction scheme~\cite{FreysoldtPRL2009}. The chemical potential difference of $\delta\mu$ is defined as 
\begin{equation}
\delta\mu = (\mu_\text{Si} - \mu_\text{C}) - (\mu_\text{Si}^\text{b} - \mu_\text{C}^\text{b}),  
\label{eq:deltamu}
\end{equation}
where $\mu_\text{Si}$ and $\mu_\text{C}$ are the chemical potentials of the Si and C atoms, respectively, in the SiC lattice obeying the $\mu_\text{SiC} = \mu_\text{Si} + \mu_\text{C}$ relation. The heat of formation ($\delta H$) of the Si-C pair in 4H SiC can be defined as 
\begin{equation}
\delta H = \mu_\text{Si}^\text{b} + \mu_\text{C}^\text{b} - \mu_\text{SiC} \text{.}
\label{eq:deltaH}
\end{equation}
Comparing Eqs.~\ref{eq:deltamu}~and~\ref{eq:deltaH} implies that $\delta \mu$ is limited by $\delta H$, i.e. under extremely C-rich condition ($\mu_\text{C} = \mu_\text{C}^\text{b}$) $\delta\mu~=~-\delta H$ while for the Si-rich limit ($\mu_\text{Si} = \mu_\text{Si}^\text{b}$)  $\delta\mu~=~\delta H$, while in the stoichiometric case $\delta \mu = 0$. The corresponding values calculated by HSE06 functional are listed in Table~\ref{tab:HSEmu}.

\begin{table}[t]
\begin{ruledtabular}
\caption {Heat of formation ($\delta H$) of 4H SiC and chemical potential values for Si atom in bulk Si ($\mu_\text{Si}^\text{b}$), for C atom in diamond ($\mu_\text{C}^\text{b}$) and for the Si-C pair in 4H SiC ($\mu_\text{SiC}$) calculated by HSE06 functional.}
\label{tab:HSEmu}
\begin{tabular}{lcccc}
{Functional}
& {$\mu_\text{Si}^\text{b} $ (eV)}
& {$\mu_\text{C}^\text{b} $ (eV)}
& {$\mu_\text{SiC}$ (eV)}
& {$\delta H$ (eV)}\\
\hline
{PBE}  & -5.42 & -9.10 & -15.06 & -0.54 \\
\end{tabular}
\end{ruledtabular}
\end{table}

Since our assumption regarding the defect models is that the V$_\text{Si}$ defect is in its single negative charge state, it is reasonable to calculate the formation and binding energies within the region between the $(0/-)$ and $(-/2-)$ adiabatic charge transition levels of the V$_\text{Si}$ defects. The adiabatic charge transition levels can be derived from Eq.~\ref{eq:formE} as follows,
\begin{equation}
E_{q+1/q} = E_{\text{tot}}^q - E_{\text{tot}}^{q+1} + \Delta V(q) - \Delta V(q+1).
\label{eq:CHLs}
\end{equation}

\subsection{\label{subsec:ZFSmethod}Zero-field splitting parameters for $S$ = 3/2 systems}
 
Zero-field splitting (ZFS) of energy levels manifests in systems with $S \geq$ 1 spin. The corresponding Hamiltonian reads as

\begin{equation}
\hat{H}_{\text{ss}} = \mathbf{\hat{S}}^\top\mathbf{D}\mathbf{\hat{S}},
\label{eq:ZFS}
\end{equation}
where $\mathbf{\hat{S}} = \sum_i \mathbf{\hat{S}}_i$ is the total spin operator obtained as the superposition of the $\mathbf{\hat{S}}_i$ one particle spin operators. In Eq.~\ref{eq:ZFS} the ZFS tensor in represented by $\mathbf{D}$ and using its diagonalized form the the spin-spin Hamiltonian takes the form of

\begin{equation}
\hat{H}_{\text{ss}} = D_{xx}\hat{S_x}^2 + D_{yy}\hat{S_y}^2 + D_{zz}\hat{S_z}^2,
\label{eq:diagZFS}
\end{equation}
where $D_{ij}$ are elements of $\mathbf{D}$-tensor and $\hat{S_x}, \hat{S_y}$ and $\hat{S_z}$ are the components of $\mathbf{\hat{S}}$ in the $x, y$ and $z$ directions, respectively.

Introducing the $D$ and $E$ ZFS parameters, i.e.~the respective axial and orthorombic components, eigenvalue of ${\hat{H}}_{\text{ss}}$ ($E_{\text{ss}}$) can be written as

\begin{equation}
E_{\text{ss}} = D \Bigg(m_S^2 - \frac{S(S+1)}{3} \Bigg) + E(S_x^2+S_y^2),
\label{eq:parZFS}
\end{equation}
where we use that the eigenvalue of $\mathbf{\hat{S}^2}$ is $S(S+1)$ with $S$ being the eigenvalue of $\mathbf{\hat{S}}$ and the eigenvalues of $\hat{S}_x, \hat{S}_y$ and $\hat{S}_z$ are $S_x, S_y$ and $m_S$, respectively. The ZFS parameters can be expressed as $D = 3D_{zz}/2$ and $E = (D_{yy}-D_{xx})/2$. Under C$_\text{3v}$ symmetry the $\mathbf{D}$-tensor contains two principal values, i.e. $D_{xx} = D_{yy}$ and $D_{zz}$ and hence the orthorombic term is zero, $E = 0$, simplifying Eq.~\ref{eq:parZFS} to 

\begin{equation}
E_{\text{ss}} =  D \Bigg(m_S^2 - \frac{S(S+1)}{3} \Bigg).
\label{eq:C3vZFS}
\end{equation}

In our calculations, the spin quantization axis, i.e.~the $z$-axis is aligned with the $c$-axis and $S = \frac{3}{2}$ for the investigated defects implying $m_S = \{\pm\frac{3}{2}, \pm\frac{1}{2}\}$. The corresponding eigenvalues are $-D$ for $m_S = \pm 1/2$ and $D$ for $m_s = \pm 3/2$ states implying the splitting of $2D$ between the spin levels.

We calculated the matrix elements of the $\mathbf{D}$-tensor as implemented by Iv\'ady.~\textit{et al.}~\cite{IvadyPRB2014ZFS}. The corresponding elements can be calculated as~\cite{IvadyPRB2014ZFS, RaysonPRB2008}

\begin{widetext}
\begin{equation}
D_{kl} = \frac{\mu_0\mu_\text{B}^2g_0^2}{S(2S-1)}\nsum[2]_{i<j}^{n} \chi_{ij} \iint \rho^{(2)}(\mathbf{r}_1, \mathbf{r}_2) \Big( \frac{|\mathbf{r}_2-\mathbf{r}_1|^2\delta_{kl}-3(\mathbf{r}_2-\mathbf{r}_1)_k \cdot (\mathbf{r}_2-\mathbf{r}_1)_l}{|\mathbf{r}_2-\mathbf{r}_1|^5} \Big) d\mathbf{r}_1d\mathbf{r}_2,
\label{eq:Delem}
\end{equation}
\end{widetext}
where the constants of $\mu_0$, $\mu_\text{B}$ and $g_0$ are the vacuum permeability, the Bohr-magneton and the free electron $g$-factor, respectively. The summation goes over every pairs of occupied Kohn-Sham states and $\chi_{kl}$ is +1 for parallel and $-1$ for antiparallel spins. The integral generates the expectation value of the dipole momentum operator on the two-particle electron density of $\rho^{(2)}(\mathbf{r}_1, \mathbf{r}_2)$ depending on the positions of the two electrons, $\mathbf{r}_1$ and $\mathbf{r}_2$.

\section{\label{sec:resdisc}Results and discussion}

In this Section we provide our numerical results for all defect models depicted on Fig.~\ref{fig:defstructs}. In this section, we report our results about the electronic structure (Sec.~\ref{subsec:elstruct}), defect formation (Sec.~\ref{subsec:defform}) and the magneto-optical parameters (Sec.~\ref{subsec:magneto}).

\subsection{\label{subsec:elstruct} Electronic structure}
\begin{figure*} [t]\centering
\includegraphics[width=0.8\textwidth]{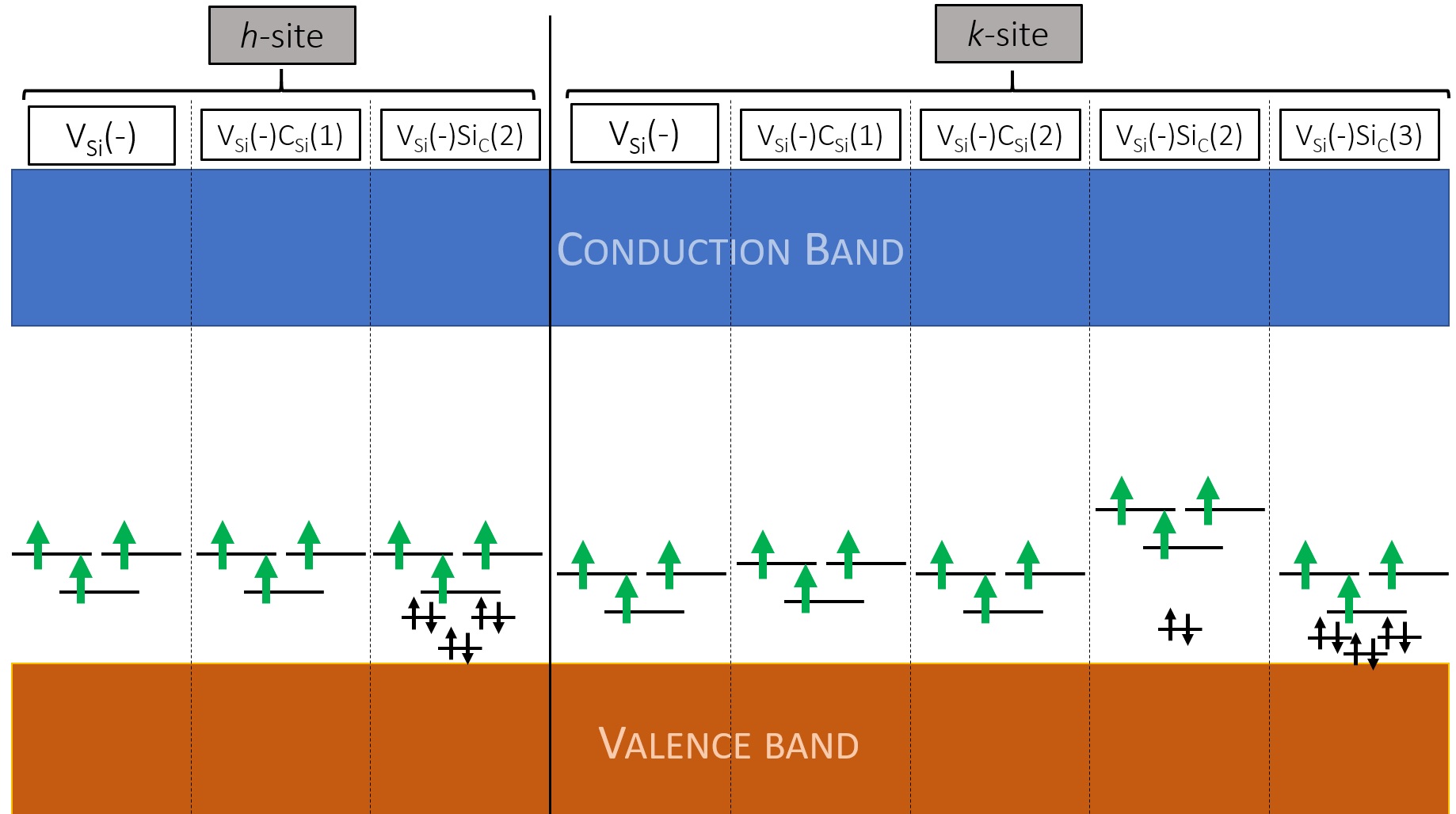}
\caption{Electronic structure of all investigated defect models. Notations of defects are aligned with those in Fig.~\ref{fig:defstructs}. Green arrows represent the electrons of V$_\text{Si}(-)$ defects residing at open orbitals, i.e.~the active space of the electronic structure. Black arrows stand for electrons at closed energy levels introduced by Si$_\text{C}(0)$. Valence and conduction bands of 4H SiC are also indicated. }
\vspace{0 pt}
\label{elstructs}
\end{figure*}

\begin{figure} [b]
\centering
\includegraphics[width=0.5\textwidth]{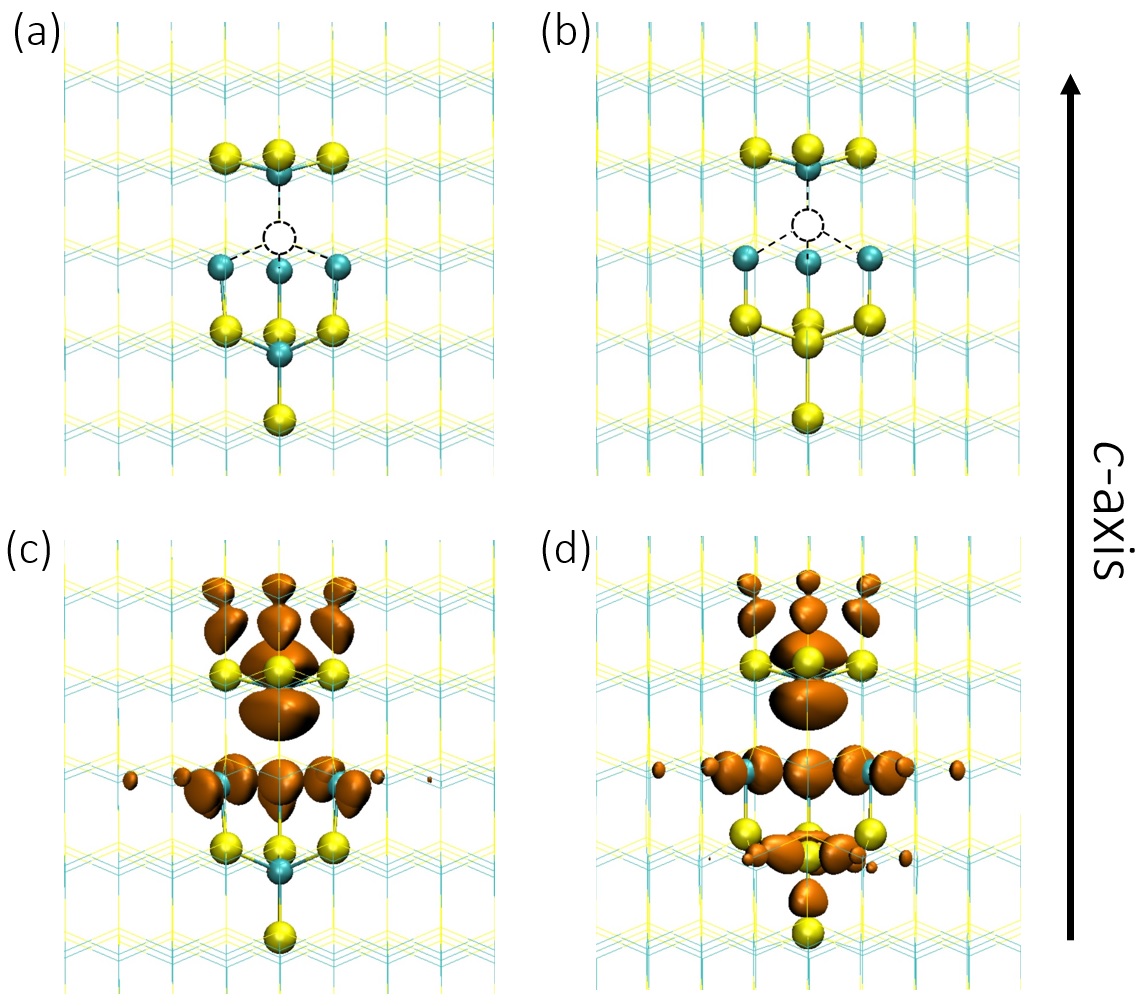}
\caption{Defect structure of (a) V$_\text{Si}(-)$-$k$ and (b) V$_\text{Si}(-)$-Si$_\text{C}(0)$-$k$ and the corresponding spin densities (orange lobes) generated by using the same isovalues, (c) and (d), respectively. The supercell structure is shown in ortographic view and the lattice is represented by a wire structure. In the core of the defects Si and C atoms are represented by yellow and cyan balls, respectively, while dashed balls stand for the V$_\text{Si}$.  }
\vspace{0 pt}
\label{fig:ksic2spindens}
\end{figure}

Electronic structures for the investigated defects complexes are illustrated in Fig.~\ref{elstructs}.
Accordingly, isolated V$_\text{Si}(-)$ defects at both $h$ and $k$ sites introduce $a_1$ and $e$ levels into the band gap of 4H SiC and both of them are half-filled establishing high-spin state of $S = 3/2$ as already known from previous studies. Introducing C$_\text{Si}(0)$ to the system the electronic structure remains very similar regardless to the neighboring order, i.e. the distance between V$_\text{Si}(-)$ and C$_\text{Si}(0)$. In particular, changing in the KS energy levels with respect to those of  V$_\text{Si}(-)$ defects is $\leq$~0.05~eV. On the other hand, when Si$_\text{C}(0)$ is included additional --- fully occupied --- KS level(s) appear in the band gap near to the valence band minimum (VBM). 
\begin{figure} [b]
\centering
\includegraphics[width=0.4\textwidth]{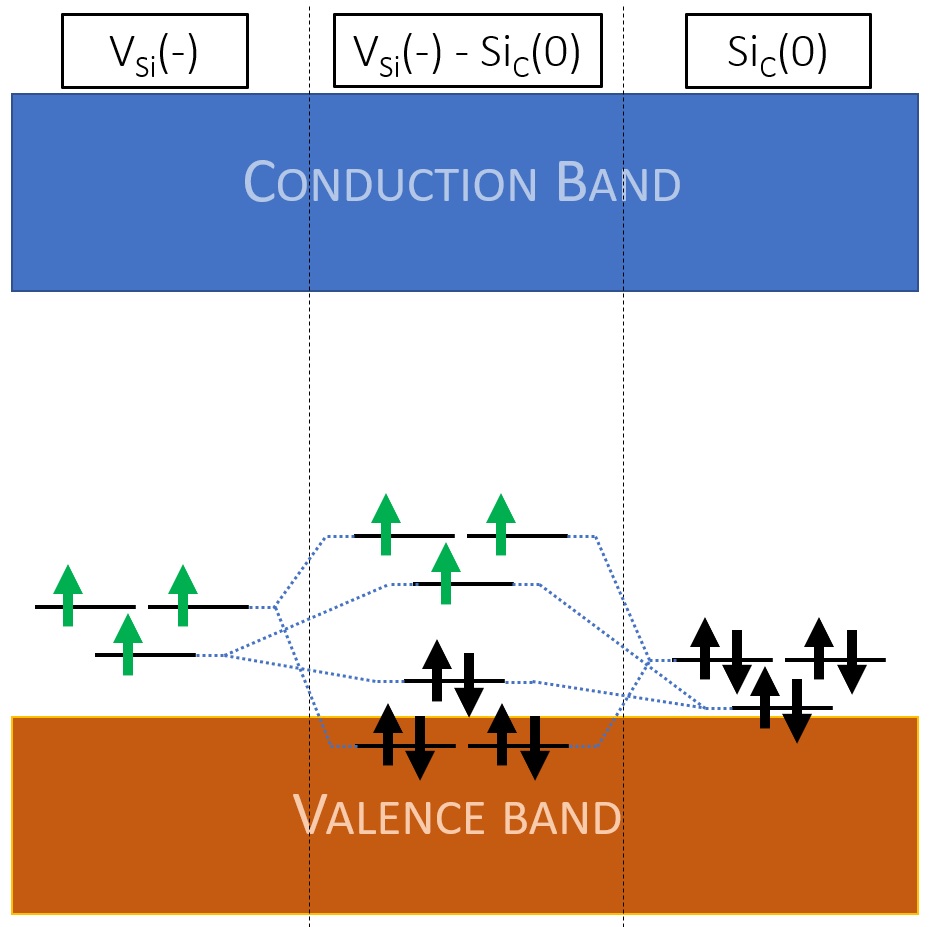}
\caption{Combination of the orbitals of V$_\text{Si}(-)$-$k$ and Si$_\text{C}(0)$-$k$ yielding mixed states, particularly, a fully occupied $e$ and $a_1$ states, where the $e$ level falls into the valence band (VB). The fully occupied $a_1$ state and further, half-occupied $a_1$ and $e$ states lie in the band gap. Green arrows stand for the unpaired electrons establishing the spin density [orange lobes in Fig~\ref{fig:ksic2spindens}~(d)], whereas black arrows represent the paired electrons. }
\vspace{0 pt}
\label{fig:defmoldiag}
\end{figure}
Electronic structure of V$_\text{Si}(-)$-$h$ with second neighboring Si$_\text{C}(0)$ is akin to that of V$_\text{Si}(-)$-$k$ with third neighboring Si$_\text{C}(0)$, i.e. a fully occupied $a_1$ and $e$ levels appear besides the half-occupied V$_\text{Si}(-)$ orbitals. Positions of the half-occupied levels are close to those of the isolated V$_\text{Si}(-)$, whereas the levels of the fully occupied states are very similar to those of Si$_\text{C}(0)$. This implies that V$_\text{Si}(-)$-$h$ and Si$_\text{C}(0)$ establish their electronic structures almost independently, i.e., their interaction is negligible. In contrast, for V$_\text{Si}(-)$-$k$ with second neighbor Si$_\text{C}(0)$ change in the electronic structure with respect either the isolated V$_\text{Si}(-)$-$k$ and the isolated Si$_\text{C}(0)$ defects is significant. Especially, KS orbitals of V$_\text{Si}(-)$ are pushed up by about 0.3~eV whereas only one fully occupied $a_1$ level appear higher by about 0.1~eV than that of the isolated Si$_\text{C}(0)$-$k$, furthermore, no fully occupied $e$ level emerges in the band gap. This significant effect implies that V$_\text{Si}(-)$-$k$ and Si$_\text{C}(0)$-$k$ cannot be treated individually in this case, i.e. Si$_\text{C}(0)$-$k$ may not be only perturbation on V$_\text{Si}(-)$-$k$. In order to study the interaction between V$_\text{Si}(-)$-$k$ and Si$_\text{C}(0)$-$k$ in detail we derive the corresponding defect-molecule diagram as depicted in Fig.~\ref{fig:defmoldiag}. Accordingly, KS states in the band gap are built up as the combination of the symmetry-linked orbitals of V$_\text{Si}(-)$-$k$ and Si$_\text{C}(0)$-$k$. i.e. these states belong to both defects. Localization of the corresponding spin density is illustrated in Fig.~\ref{fig:ksic2spindens} where we show also the spin density of the isolated V$_\text{Si}(-)$-$k$ for comparison. 
In particular, the spin density corresponding to the V$_\text{Si}(-)$-Si$_\text{C}(0)$-$k$ complex is significantly expanded to the Si$_\text{C}(0)$-$k$ [cf.~Fig.~\ref{fig:ksic2spindens}~(d)]. In contrast, for the isolated V$_\text{Si}(-)$-$k$ [cf.~Fig.~\ref{fig:ksic2spindens}~(c)] there is a negligible contribution of the spin density on the C atom that is replaced by the Si atom in the V$_\text{Si}(-)$-Si$_\text{C}(0)$-$k$ defect comlplex. As a result, the corresponding $D$ constant may significantly differ from the observed values listed in Table~\ref{tab:expDconst}. Indeed, we obtain the $D$ constant value of 173.87~MHz for this defect complex that is one order of magnitude larger than the experimental values for the unknown EPR centers. This further supports that Si$_\text{C}(0)$-$k$ at the second neighbor C-site of V$_\text{Si}(-)$ cannot be treated as only a weak perturbation on the electronic structure of V$_\text{Si}(-)$ and hence we exclude this defect model from further investigations in the context.

\subsection{\label{subsec:defform} Defect formation}

In order to determine the binding energies of the $\text{V}_\text{Si}(-)\text{-}X$ complexes, we calculated the formation energies as functions of the Fermi level using the introduced defect models reported in Sec.~\ref{sec:defmod}. Since EPR signatures of V$_\text{Si}(-)$-$h/k$ are observed in the corresponding experimental spectra~\cite{SonIOP2019}, we calculated the formation and binding energies within the Fermi level region, where V$_\text{Si}(-)$-$h/k$ is stable, i.e. between the $(0/-)$ and $(-/2-)$ charge transition levels of V$_\text{Si}(-)$-$h/k$. Furthermore, we also assume that C$_\text{Si}$ and Si$_\text{C}$ are in their neutral charge state, i.e. $q = 0$ within this region implying the last two terms in Eq.~\ref{eq:formE} to be vanished for these defects. 

Charge transition levels of V$_\text{Si}(-)$-$h/k$ calculated by HSE06 functional are reported in Table~\ref{tab:VSiCTLHSE}. We note that similar values are reported in Refs.~\onlinecite{TorpoJP2001, KobayashiJAP2019, DavidJAP2004, HornosMatSci2011}. 
\begin{table}[b]
\begin{ruledtabular}
\caption {\small Values for the $(0/-)$ and $(-/2-)$ charge transition levels of V$_\text{Si}(-)$-$h/k$ defects in 4H SiC as calculated by HSE06 functional. All values are referenced to the valence band maximum.}
\label{tab:VSiCTLHSE}
\begin{tabular}{lccc}
{Functional}
& {Defect} 
& {$E_{(0/-)}$ (eV)}
& {$E_{(-/2-)}$} (eV)\\
\hline
\multirow{2}{*}{HSE06} & V$_\text{Si}(-)$-$h$ & 1.29 & 2.59 \\
& V$_\text{Si}(-)$-$k$ & 1.26 & 2.47\\
\end{tabular}
\end{ruledtabular}
\end{table}

\begin{figure} [t]
\centering
\includegraphics[width=0.45\textwidth]{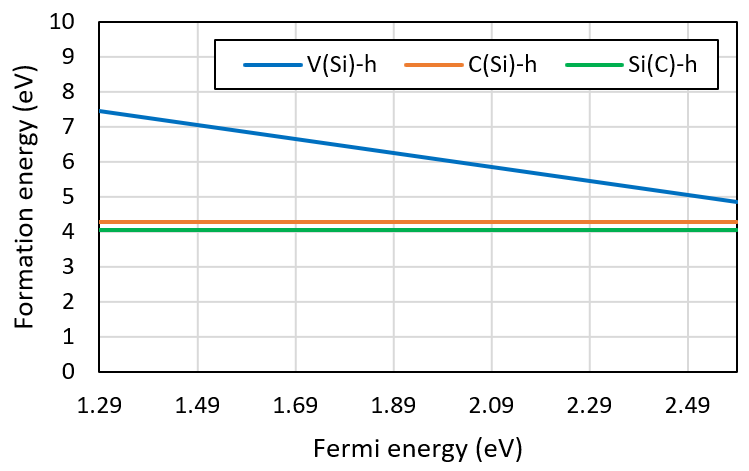}
\caption{Formation energies of isolated V$_\text{Si}$, C$_\text{Si}$ and Si$_\text{C}$ defects under stoichiometric conditions calculated by HSE06 functional in the Fermi-level region between the $(0/-)$ and $(-/2-)$ charge transition levels of V$_\text{Si}$-$h/k$. }
\vspace{20 pt}
\label{fig:formeisoHSE}
\end{figure}

Formation energies as a function of the Fermi-level for the isolated defect species under stoichiometric conditions are reported in Table~\ref{tab:formebindeHSE} and depicted for the hexagonal defects in Fig.~\ref{fig:formeisoHSE}. We note that similar values were obtained from previous local spin density approximation DFT calculations~\cite{TorpoJP2001, KobayashiJAP2019}. Based on the numerical results V$_\text{Si}(-)$ defects exhibit the largest formation energy across the investigated Fermi-energy region. Formation energies of V$_\text{Si}(-)$-C$_\text{Si}$ defect complexes are higher by about 0.1~eV than those of V$_\text{Si}(-)$-Si$_\text{C}$ defect complexes.

Based on the calculated formation energies we calculated the binding energies of the corresponding V$_\text{Si}(-)$-$X$ defect complexes by using Eq.~\ref{eq:binde}. Since both $\delta\mu$ and $E_\text{F}$ are cancelled out, all defect complexes exhibit constant binding energy values across the investigated $E_\text{F}$ region and also insensitive to the chemical environment.  The obtained values are listed in Tab.~\ref{tab:formebindeHSE}. Generally, all defect models exhibit small binding energies supporting that the presence of the $X$ defect species are indeed a perturbation for V$_\text{Si}(-)$ establishing only weakly bound complexes. Since the binding energy values are close to each other for the different V$_\text{Si}(-)$-$X$ complexes, they may appear in the same order of magnitude concentration during the preparation of a 4H SiC sample.

\begin{table}[t]
\begin{ruledtabular}
\caption{Formation ($E_\text{form}$) and binding energies ($E_\text{b}$) of the V$_\text{Si}(-)$-$X$ defect complexes calculated HSE06 functional in the Fermi-level ($E_\text{F}$) region between the  $(0/-)$ and $(-/2-)$ charge transition levels of V$_\text{Si}$-$h/k$ under stoichiometric conditions.}
\label{tab:formebindeHSE}
\begin{tabular}{lcc}
{Defect}
 & $E_\text{form}$ (eV) & $E_\text{b}$ (eV)\\
\hline

V$_\text{Si}(-)$-$h$ & 7.45 - $E_\text{F}$ & -  \\

V$_\text{Si}(-)$-$k$ &  7.47 - $E_\text{F}$ & -  \\

C$_\text{Si}$-$h$  & 4.27 & - \\

C$_\text{Si}$-$k$ & 4.23 & -\\

Si$_\text{C}$-$h$ & 4.06 & -\\

Si$_\text{C}$-$k$ & 4.03 & -\\

\hline

V$_\text{Si}(-)$-C$_\text{Si}$-$h$  & 11.49 - $E_\text{F}$ & 0.23 \\

V$_\text{Si}(-)$-Si$_\text{C}$-$h$  & 11.38 - $E_\text{F}$& 0.14 \\

V$_\text{Si}(-)$-C$_\text{Si}$-$k$  & 11.46 - $E_\text{F}$ & 0.24 \\

V$_\text{Si}(-)$-C$_\text{Si}(2)$-$k$ & 11.46-$E_\text{F}$ & 0.24 \\

V$_\text{Si}(-)$-Si$_\text{C}(3)$-$k$ & 11.34 - $E_\text{F}$ & 0.15 \\

\end{tabular}
\end{ruledtabular}
\end{table}

\subsection{\label{subsec:magneto} Magneto-optical signatures}

All V$_\text{Si}(-)$-$X$ defect complexes introduces similar spin-3/2 electronic structure to that of the isolated V$_\text{Si}(-)$ defects, i.e. $^4A_2$ ground state. The paramagnetic ground state makes them EPR active centers and thus their signals may appear besides that of the isolated V$_\text{Si}(-)$ defects. Indeed, several signals have been observed recently labeled as T$_\text{V1b}$, T$_\text{V2b}$, R$_1$ and R$_2$~\cite{SormanPRB2000, MizuochiPRB2002, MizuochiPRB2003, MizuochiPRB2005, JanzenPhysB2009, SonIOP2019} exhibiting similar spin properties to those of the T$_\text{V1a}$ and T$_\text{V2a}$ centers previously assigned to V$_\text{Si}(-)$-$h$ and V$_\text{Si}(-)$-$k$, respectively~\cite{IvadyPRB2017_1, DavidssonAPL2019}. Here we note that the orthorombic parameter of the ZFS, i.e. the $E$ constant is zero for all the experimentally reported signals indicating that the centers exhibit C$_\text{3v}$ symmetry. The corresponding experimental ZFS $D$-constants are listed in Table~\ref{tab:expDconst} while in Table~\ref{tab:Dconst} we report our numerical values obtained by means of PBE functional applied on 576-atom and 1536-atom supercells.
\begin{table}[t]
\begin{ruledtabular}
\caption {\small Experimental values for ZFS $D$ constants of the EPR signals reported in Ref.~\cite{SonIOP2019}. }
\label{tab:expDconst}
\begin{tabular}{lcccccc}
{Signal} & T$_\text{V1a}$ & T$_\text{V2a}$ & T$_\text{V1b}$ & T$_\text{V2b}$ & R$_\text{1}$ & R$_\text{2}$ \\
\hline
{$D_\text{exp}$ (MHz)} & 2.50 & 35.0 & 32.6 & 20.0 & 2.24 & 39.4 \\
\end{tabular}
\end{ruledtabular}

\end{table}
\begin{table}[t]
\begin{ruledtabular}
\caption {\small Calculated values for ZFS $D$ constants of the isolated V$_\text{Si}(-)$ defects and the V$_\text{Si}(-)$-$X$ defect models. The calculated values as obtained by PBE functional on 576-atom and 1536-atom supercells are denoted by $D^{576}_\text{PBE}$ and $D^{1536}_\text{PBE}$, respectively.}
\label{tab:Dconst}
\begin{tabular}{lcc}
{Defect} 
& {$D^{576}_\text{PBE}$ (MHz)}
& {$D^{1536}_\text{PBE}$ (MHz)}\\
\hline
V$_\text{Si}(-)$-$h$ & 18.01 & 15.23 \\
V$_\text{Si}(-)$-C$_\text{Si}$-$h$ & 25.56 & 23.03 \\
V$_\text{Si}(-)$-Si$_\text{C}$-$h$ & 18.75 & 21.34 \\

V$_\text{Si}(-)$-$k$  & 24.99 & 34.19 \\
V$_\text{Si}(-)$-C$_\text{Si}$-$k$ & 2.79 & 2.29\\
V$_\text{Si}(-)$-C$_\text{Si}(2)$-$k$ & 23.92 & 29.41\\
V$_\text{Si}(-)$-Si$_\text{C}(3)$-$k$ & 33.57 & 33.55\\
\end{tabular}
\end{ruledtabular}
\end{table}
Accordingly, numerical values for the isolated V$_\text{Si}(-)$ agree with the previously reported ones also yielded by DFT calculations~\cite{IvadyPRB2017_1, DavidssonAPL2019}. Regarding the V$_\text{Si}(-)$-$X$ defects, all calculated $D$ values fall into the 18-34~MHz region for both supercell calculations except for V$_\text{Si}(-)$-C$_\text{Si}$-$k$ defect, where the corresponding $D$ constant is one order of magnitude lower. This might be consequence that the distance between the V$_\text{Si}(-)$ and C$_\text{Si}$ defect species is the shortest and hence the presence of C$_\text{Si}$ affects the most the spin density of V$_\text{Si}(-)$ among the other defect complexes (see Fig.~\ref{fig:defstructs}). Here we note that distance between V$_\text{Si}(-)$ and Si$_\text{C}$ antisite in the V$_\text{Si}(-)$-Si$_\text{C}(2)$-$k$ defect model is even shorter, however, it is already excluded from our recent scope (see in Sec.~\ref{subsec:elstruct}). This is also the case for the experimental signals where all $D$ constants of the unknown centers take place within the 20-40~MHz interval except for the R$_1$ center exhibiting a one order of magnitude lower $D$ constant. As a consequence we attribute the R$_1$ signal to the V$_\text{Si}(-)$-C$_\text{Si}$-$k$ defect. We cannot unambigously identify the other unknown EPR centers based on solely the corresponding $D$ constants.

\begin{table}[b]
\begin{ruledtabular}
\caption{Positions of ZPLs for the investigated V$_\text{Si}(-)$-$X$ defect models calculated by means of PBE functional. We also report the values yielded by HSE06 functional for the isolated V$_\text{Si}(-)$ defects. The ZPL values in the HSE06* column for the defect complexes were obtained by correcting the corresponding PBE result with the difference of the isolated V$_\text{Si}(-)$-related PBE and HSE06 ZPL values as an estimate.}
\label{tab:ZPL}
\begin{tabular}{lcc}
{Defect} 
& {$E_\text{ZPL}^{\text{PBE}}$~(eV)}
& {$E_\text{ZPL}^{\text{HSE06*}}$~(eV)}\\
\hline
V$_\text{Si}(-)$-$h$ & 1.273 & 1.450 \\
V$_\text{Si}(-)$-C$_\text{Si}$-$h$ & 1.267 & 1.444 \\
V$_\text{Si}(-)$-Si$_\text{C}$-$h$ & 1.107 & 1.284\\
\hline
V$_\text{Si}(-)$-$k$  & 1.198 & 1.385 \\
V$_\text{Si}(-)$-C$_\text{Si}$-$k$ & 1.345 & 1.532\\
V$_\text{Si}(-)$-C$_\text{Si}(2)$-$k$& 1.182 & 1.369 \\
V$_\text{Si}(-)$-Si$_\text{C}(3)$-$k$& 1.280 & 1.467\\
\end{tabular}
\end{ruledtabular}
\end{table}

Optical signature of point defects provide additional fingerprints in defect identification. Although, fluorescence centers associated with these EPR centers have not yet been reported, we calculated the ZPL energies for the isolated V$_\text{Si}(-)$ defects and for all investigated defect complexes as listed in Table~\ref{tab:ZPL}. For V$_\text{Si}(-)$ defects the experimental ZPL energies are 1.438~eV for the V1 (T$_\text{V1a}$ in EPR) and 1.352~eV for the V2 (T$_\text{V2a}$ in EPR) PL centers identified as V$_\text{Si}(-)$-$h$ and V$_\text{Si}(-)$-$k$, respectively~\cite{IvadyPRB2017_1, DavidssonAPL2019}. The lowest and largest ZPL energies are found for V$_\text{Si}(-)$-Si$_\text{C}$-$h$ and V$_\text{Si}(-)$-C$_\text{Si}$-$k$ defects, respectively. The calculated optical signals may be detected in future experiments that may lead to the identification of the considered defect complexes.

\begin{figure}[h]
\centering
\includegraphics[width=0.4\textwidth]{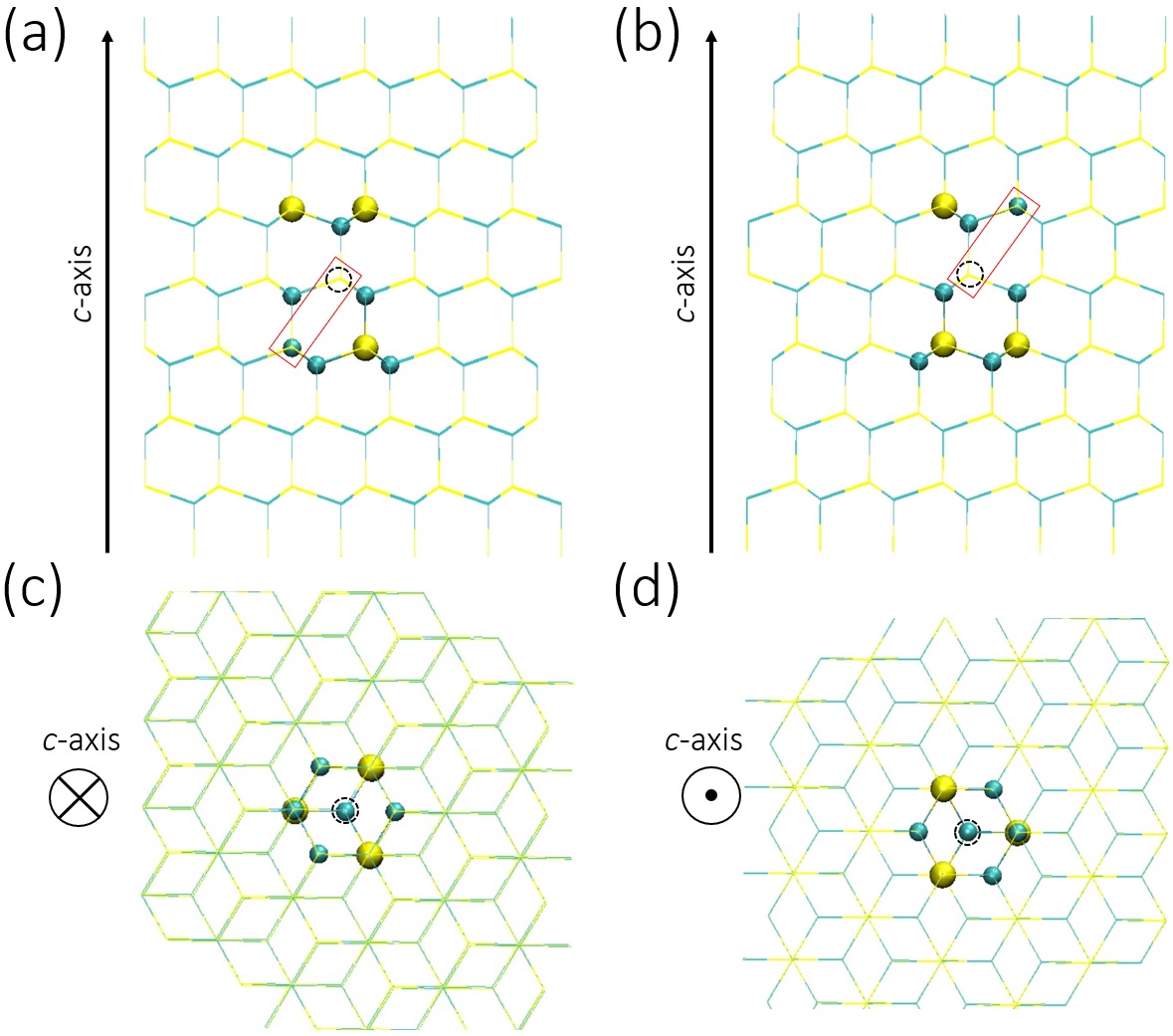}
\caption{Defect structures of the investigated basal (a)-(c) V$_\text{Si}(-)$-C$_\text{Si}$-$hk^{\text{(a)}}$ and (b)-(d) V$_\text{Si}(-)$-C$_\text{Si}$-$hk^{\text{(b)}}$ defect configurations. The supercell structure is shown in ortographic view and the lattice is represented by a wire structure. In the core of the defects Si and C atoms are represented by yellow and cyan balls, respectively, while dashed balls stand for the V$_\text{Si}$. We indicate the $c$-axis for all views. }
\vspace{20 pt}
\label{fig:basVSiCSi}
\end{figure}

\begin{figure*}[h]
\centering
\includegraphics[width=0.8\textwidth]{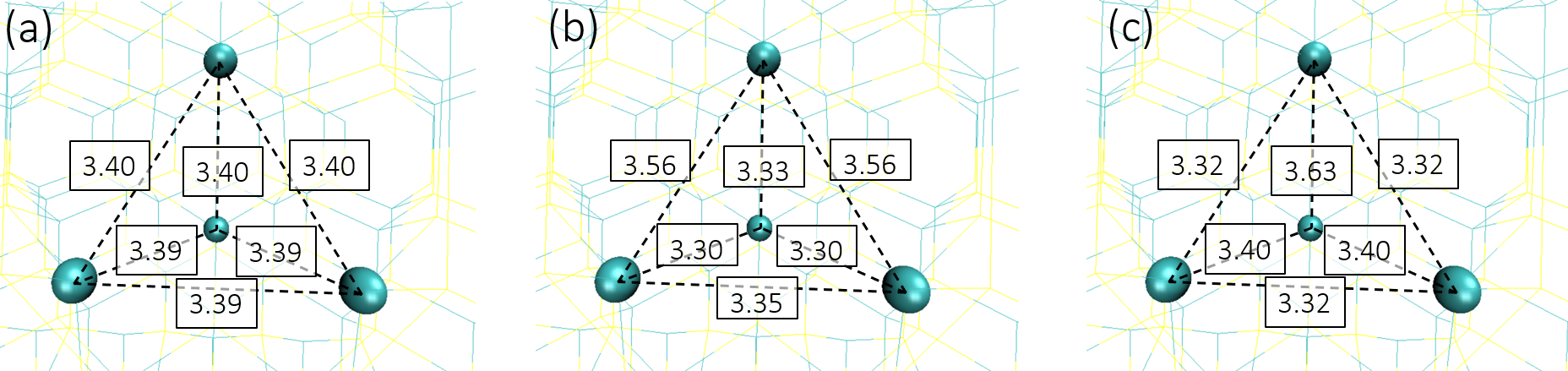}
\caption{Distortion of the tetrahedron formed by the 4 $\times$ C neighboring the V$_\text{Si}$ for (a) isolated V$_\text{Si}(-)$-$h$, (b) V$_\text{Si}(-)$-C$_\text{Si}$-$hk^{\text{(a)}}$ and (c) V$_\text{Si}(-)$-C$_\text{Si}$-$hk^{\text{(b)}}$. The 4 $\times$ C nuclei is represented by cyan balls, while the rest of the supercell is shown as wire structure. The bond lengths are given in $\AA$ units.}
\vspace{20 pt}
\label{fig:vsi_deform}
\end{figure*}

\begin{figure}[h!]
\centering
\includegraphics[width=0.45\textwidth]{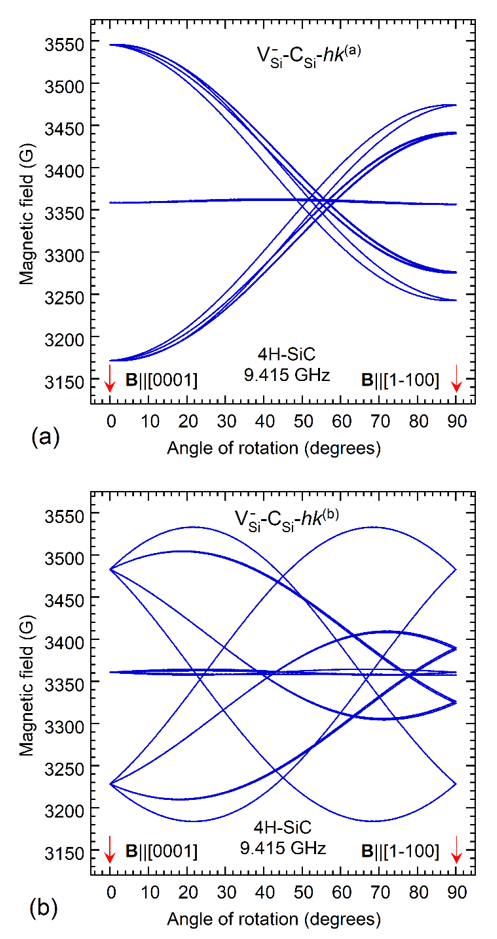}
\caption{\small Angular dependencies of the two represantive V$_\text{Si}$-related centers in 4H SiC with C$_\text{1h}$ symmetry (a) V$_\text{Si}$-C$_\text{Si}$-$hk^\text{(a)}$ and (b) V$_\text{Si}$-C$_\text{Si}$-$hk^\text{(b)}$ with the magnetic field rotating in the ($11\overline{2}0$) plane and the microwave frequency of $9.415$~GHz. In the simulations, the $g$-value of the negative Si vacancy ($g = 2.0029$) is assumed for these $S=3/2$ centers. The fine-structure parameters for the centers are $D^\text{(a)} = -226.26$~MHz, $E^\text{(a)} = -20.66$~MHz, and $D^\text{(b)} = -295.64$~MHz, $E^\text{(b)} = -59.36$~MHz are from the calculations of the ZFS tensor where the direction of the axial component has an angle of about 0.8$^\circ$ and 70.6$^\circ$, respectively, with the $c$-axis of 4H SiC.}
\vspace{20 pt}
\label{fig:basEPR}
\end{figure}

\subsection{\label{subsec:EPR_C1h} EPR of defect complexes exhibiting C$_\text{1h}$ symmetry}

Axial V$_\text{Si}(-)$-$X$ pair defects have been investigated so far but formation of basal defect configurations exhibiting C$_\text{1h}$ symmetry is also possible during the cascade process of ion collisions induced by irradiation. Indeed, in the corresponding EPR spectrum~\cite{SonIOP2019} several signals of basal centers have been observed. However, resolving these signals can be extremely challenging arising from the emerging orthrombic ZFS parameter $E$ with complex angle dependence of the spectrum as a function of the direction of the external magnetic fields, and because of the various overlapping EPR signals caused by the different configurations of pair defects. 

For demonstrating the complexity of the problem, we consider here two representative basal plane V$_\text{Si}(-)$-C$_\text{Si}$ configurations. These configurations are obtained by placing C$_\text{Si}$ in the second neighbor site with respect to V$_\text{Si}-h$. There are two possible configurations that are inequivalent by symmetry: (a)  C$_\text{Si}$ is the neighbor of $C$ dangling bond residing off the symmetry axis of the defect [see Fig.~\ref{fig:basVSiCSi}(a)] and (b) C$_\text{Si}$ is the neighbor of C dangling bond residing in the symmetry axis of the defect [see Fig.~\ref{fig:basVSiCSi}(b)]. These configurations are labeled as $_\text{Si}(-)$-C$_\text{Si}$-$hk^{\text{(a)}}$ and V$_\text{Si}(-)$-C$_\text{Si}$-$hk^{\text{(b)}}$, respectively. Since the C antisite is placed in the second nearest neighbor site of V$_\text{Si}$ the corresponding geometries are severely distorted as depicted in Fig.~\ref{fig:vsi_deform}. The large difference between the geometry of the isolated V$_\text{Si}$ [see Fig.~\ref{fig:vsi_deform}~(a)] and that of the investigated basal defects [see Fig.~\ref{fig:vsi_deform}~(b)~and~(c)] may imply similarly significant difference between the corresponding spin densities governing the ZFS $D$ and $E$ parameters. In this way, the ZFS parameters of the basal V$_\text{Si}$-related defects may highly deviates from those of the isolated V$_\text{Si}(-)$-$h$. Indeed, the corresponding calculated ZFS parameters are $D^{\text{(a)}}=-262.26$~MHz $E^{\text{(a)}}=-20.66$~MHz and $D^{\text{(b)}}=-295.64$~MHz $E^{\text{(b)}}=-59.36$~MHz. The angle of the principal axis of the $D$-tensor makes $\approx$0.8$^\circ$ and $\approx$70.6$^\circ$ angle with the $c$-axis, respectively. This large difference between the angles is due to the position of the C antisite, i.e. for V$_\text{Si}$-C$_\text{Si}$-$hk^\text{(a)}$ the C$_\text{Si}$ is positioned closer to the symmetry axis of V$_\text{Si}$ than that for V$_\text{Si}$-C$_\text{Si}$-$hk^\text{(b)}$ that implies a smaller deviation from the C$_\text{3v}$ symmetry in (a) configuration than that for (b) configuration. As a result principal axes of the $D$-tensor exhibit a small tilt from $c$-axis for V$_\text{Si}$-C$_\text{Si}$-$hk^\text{(a)}$. The corresponding $D$ constants are at least one order of magnitude higher than those for the reported axial V$_\text{Si}(-)$-$X$ defects (see~Table~\ref{tab:Dconst}) as a result of the C$_\text{Si}$ being closer to the V$_\text{Si}(-)$ than $X$ for the axial V$_\text{Si}$-$X$ complexes which significantly modifies the spin density matrix of V$_\text{Si}(-)$. 
 
Here we provide simulated EPR spectra of two representative basal V$_\text{Si}(-)$-C$_\text{Si}$ pair defects (see Fig.~\ref{fig:basEPR}) modeled in a 576-atom supercell. By using the calculated $D$-tensors we simulated the ZFS as a function of the angle of rotation about the ($1\overline{1}00$) and ($11\overline{2}0$) axes. Under the condition of $\mathbf{B} \parallel c$ the corresponding splittings are $1048.4$~MHz for V$_\text{Si}(-)$-C$_\text{Si}$-$hk^{\text{(a)}}$ and $713.9$~MHz for V$_\text{Si}(-)$-C$_\text{Si}$-$hk^{\text{(b)}}$. The (a) configuration shows such angular dependence in the EPR spectrum where the corresponding EPR transition energies closely grouped with each other, however, the (b) configuration exhibits rather a complicated pattern with split lines in the EPR spectrum as a result of the stronger C$_\text{1h}$ field originating from the C$_\text{Si}$. In experiments, both defects may present which are manifested in the EPR spectrum. By overlaying the two spectra results in a complex pattern that makes it extremely difficult to apply a spin Hamiltonian retrofit to distinguish these two centers. In experiments, other defect configurations produce other complex patterns with various $D$-tensors of scattering orthorombic $E$ components which makes the discrimination of the EPR transition energies associated with different defect configurations almost impossible.



\section{\label{sec:sum}Summary}

In summary, we carried out DFT calculations in order to identify the recently observed EPR centers~\cite{SonIOP2019} possibly associated with V$_\text{Si}(-)$ defects in 4H SiC. We set up the corresponding defect models that are complexes exhibiting C$_\text{3v}$ symmetry and are built up from a V$_\text{Si}(-)$ and a farther antisite, i.e. C$_\text{Si}$ or Si$_\text{C}$ along the $c$-axis in the 4H SiC lattice establishing V$_\text{Si}(-)$-$X$ complexes. We reported the electronic structures revealing that no further states appear in the band gap for $X$ = C$_\text{Si}$ while for $X$~=~Si$_\text{C}$ fully occupied levels appear below those of V$_\text{Si}(-)$. In particular, we found that the C$_\text{Si}(-)$-Si$_\text{C}$(2)-$k$ defect complex introduce a significantly different electronic structure and spin density with respect to that of the isolated V$_\text{Si}(-)$ yielding one order of magnitude larger $D$ constant than the experimentally observed values. 

We also investigated the formation of the defect complexes and found negligible variation in the corresponding binding energies implying defect formation with nearly the same concentrations for all the defects. We calculated the ZFS $D$-constants and compared them to the experimental values implying the V$_\text{Si}(-)$-C$_\text{Si}$-$k$ defect to be the origin of the R$_1$ EPR signal. Although experimental ZPLs are not available in the literature we also reported the corresponding values that --- along with future fluorescence or optically detected magnetic resonance measurements --- might contribute to unambiguous identification.

\section{\label{sec5} Acknowledgements}
The support from the \'UNKP-20-4 New National Excellence Program of the Ministry for Innovation and Technology from the source of the National Research, Development and Innovation Fund is acknowledged by A.\ Cs. We acknowledge the support from the BME IE-NAT TKP2020 grant of NKFIH, Hungary. We acknowledge the EU H2020 project QuanTELCO (Grant No.\ 862721). A.G.\ acknowledges the Hungarian National Quantum Technology Program (Grant No.\ 2017-1.2.1-NKP-2017-00001), the QuantERA project Nanospin (Grant No.\ NN127902) and the support from the Quantum Information National Laboratory from the Ministry for Innovation and Technology. N.T.S.~acknowledges the Swedish Research Council (Grant No.\ VR 2016-04068) and the Knut and Alice Wallenberg Foundation (Grant No.\ KAW 2018.0071). We acknowledge the computational sources provided by the Swedish National Infrastructure for Computing (SNIC) at National Computation Centre (NSC) partially funded by the Swedish Research Council through grant agreement No.\ 2018-05973 and the Governmental Agency for IT Development of Hungary through the project ``gallium''.

\begin{table}[t]
\begin{ruledtabular}
\caption {Heat of formation ($\delta H$) of 4H SiC and chemical potential values for Si atom in bulk Si ($\mu_\text{Si}^\text{b}$), for C atom in diamond ($\mu_\text{C}^\text{b}$) and for the Si-C pair in 4H SiC ($\mu_\text{SiC}$) calculated by PBE functional.}
\label{tab:muPBE}
\begin{tabular}{cccc}
{$\mu_\text{Si}^\text{b} $ (eV)}
& {$\mu_\text{C}^\text{b} $ (eV)}
& {$\mu_\text{SiC}$ (eV)}
& {$\delta H$ (eV)}\\
\hline
-5.42 & -9.10 & -15.06 & -0.54 \\
\end{tabular}
\end{ruledtabular}
\end{table}

\begin{table}[t]
\begin{ruledtabular}
\caption {\small Values for the $(0/-)$ and $(-/2-)$ charge transition levels of V$_\text{Si}(-)$-$h/k$ defects in 4H SiC as calculated by PBE functional. All values are referenced to the valence band maximum.}
\label{tab:VSiCTLPBE}
\begin{tabular}{ccc}
{Defect} 
& {$E_{(0/-)}$ (eV)}
& {$E_{(-/2-)}$} (eV)\\
\hline
 V$_\text{Si}(-)$-$h$ & 0.64 & 1.58 \\
 V$_\text{Si}(-)$-$k$ & 0.57 & 1.60\\
\end{tabular}
\end{ruledtabular}
\end{table}
\bigskip

\begin{table}[t]
\begin{ruledtabular}
\caption{Formation ($E_\text{form}$) and binding energies ($E_\text{b}$) of the V$_\text{Si}(-)$-$X$ defect complexes calculated PBE functional in the Fermi-level ($E_\text{F}$) region between the  $(0/-)$ and $(-/2-)$ charge transition levels of V$_\text{Si}$-$h/k$ under stoichometric condition.}
\label{tab:formebindePBE}
\begin{tabular}{lcc}
{Defect}
& $E_\text{form}$ (eV) & $E_\text{b}$ (eV) \\

\hline
V$_\text{Si}(-)$-$h$ & 7.05 - $E^\text{PBE}_\text{F}$  & -\\

V$_\text{Si}(-)$-$k$ & 7.06 - $E^\text{PBE}_\text{F}$ & - \\

C$_\text{Si}$-$h$  & 3.43& - \\

C$_\text{Si}$-$k$ & 3.39& - \\

Si$_\text{C}$-$h$ & 3.96& -\\

Si$_\text{C}$-$k$ & 4.00& - \\

\hline

V$_\text{Si}(-)$-C$_\text{Si}$-$h$ & 10.46 - $E_\text{F}$ & 0.02  \\

V$_\text{Si}(-)$-Si$_\text{C}$-$h$ & 11.08 - $E_\text{F}$& -0.07 \\

V$_\text{Si}(-)$-C$_\text{Si}$-$k$ & 10.41 - $E_\text{F}$ & 0.04 \\

V$_\text{Si}(-)$-C$_\text{Si}(2)$-$k$ & 10.41 - $E_\text{F}$ & 0.03 \\

V$_\text{Si}(-)$-Si$_\text{C}(3)$-$k$ & 11.04 - $E_\text{F}$ & 0.02 \\

\end{tabular}
\end{ruledtabular}
\end{table}

\appendix
\section{Numerical results for the formation of V$_\text{Si}(-)$-$X$ complexes by means of PBE functional}

\begin{figure} [b]
\centering
\includegraphics[width=0.45\textwidth]{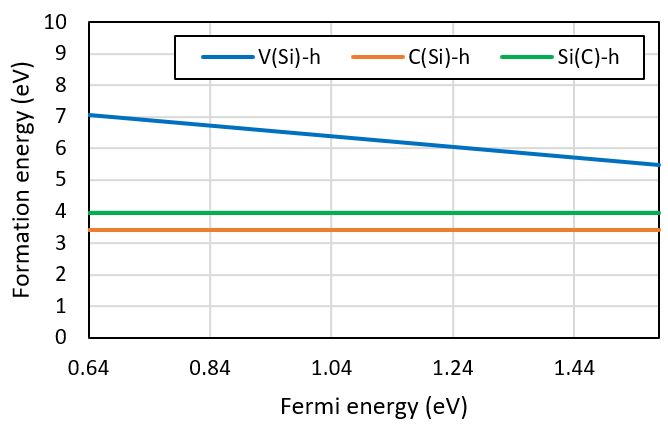}
\caption{Formation energies of isolated V$_\text{Si}$, C$_\text{Si}$ and Si$_\text{C}$ defects  under stoichiometric conditions calculated by PBE functional in the Fermi-level region between the $(0/-)$ and $(-/2-)$ charge transition levels of V$_\text{Si}$-$h/k$. }
\vspace{20 pt}
\label{fig:formeisoPBE}
\end{figure}

We calculated the formation and binding energies for the V$_\text{Si}(-)$-$X$ axial complexes also by means of the PBE~\cite{PerdewPRL1996} functional. Here we report these results for the comparison with that yielded by the HSE06 functional. In Table~\ref{tab:muPBE} we present the parameters for the formation energy calculations, i.e. for Eq.~\ref{eq:formE}.

$(0/-)$ and $(-/2-)$ charge transition levels of V$_\text{Si}$ defects --- designating the Fermi-level region for the formation and binding energies --- are presented in Table~\ref{tab:VSiCTLPBE}.

Accordingly, results yielded by PBE functionals are lower by about 0.6-0.7~eV for the $(0/-)$ level and by about 0.9-1.0~eV for $(-/2-)$ level than those obtained by the HSE06 functional listed in Table~\ref{tab:VSiCTLHSE}. On the other hand, both functionals predict that both charge transition levels lie higher for V$_\text{Si}(-)$-$h$ than those for V$_\text{Si}(-)$-$k$.

Formation energies for the V$_\text{Si}(-)$-$X$ defect complexes are listed in Table~\ref{tab:formebindePBE} and depicted in Fig.~\ref{fig:formeisoPBE}. Values for V$_\text{Si}(-)$-C$_\text{Si}$ defect models are lower by about 0.6~eV than those for the V$_\text{Si}(-)$-Si$_\text{C}$ defect complexes. Generally, all formation energy values calculated by means of HSE06 functional (cf.\ Table~\ref{tab:formebindeHSE}) are higher than the PBE ones. However, HSE06 functional predicts one order of magnitude larger binding energies --- and thus higher stability --- for all defect models than those calculated by PBE functional.



%

\end{document}